\documentclass[conference]{IEEEtran}
\IEEEoverridecommandlockouts
\usepackage{cite}
\usepackage[colorinlistoftodos,prependcaption]{todonotes}
\usepackage{amsmath,amssymb,amsfonts}
\usepackage{graphicx}
\usepackage{textcomp}
\usepackage{xcolor}
\usepackage{gensymb}
\usepackage{hyperref}
\usepackage{breakurl}
\usepackage{url}
\def\BibTeX{{\rm B\kern-.05em{\sc i\kern-.025em b}\kern-.08em
    T\kern-.1667em\lower.7ex\hbox{E}\kern-.125emX}}
\begin{document}

\title{Offloading Execution from Edge to Cloud:\\ a Dynamic Node-RED Based Approach
\thanks{This research has been supported by the EU under the H2020 AGILE
(Adaptive Gateways for dIverse muLtiple Environments), grant
agreement number H2020-688088.
\newline\copyright2018 IEEE}
}

    \author{\IEEEauthorblockN{Rom\'an Sosa}
\IEEEauthorblockA{\textit{Research \& Innovation Group} \\
\textit{ATOS}\\
Tenerife, Spain \\
roman.sosa@atos.net}
\and
\IEEEauthorblockN{Csaba Kiraly}
\IEEEauthorblockA{\textit{OpenIoT Research Unit} \\
\textit{FBK CREATE-NET}\\
Trento, Italy \\
kiraly@fbk.eu}
\and
\IEEEauthorblockN{Juan D. Parra Rodriguez}
\IEEEauthorblockA{\textit{IT-Security Group} \\
\textit{ University of Passau}\\
Passau, Germany \\
dp@sec.uni-passau.de}
}

\maketitle

\begin{abstract}
Fog computing enables use cases where data produced in end devices are stored, processed, and acted on directly at the edges of the network, yet computation can be offloaded to more powerful instances through the edge to cloud continuum. Such offloading mechanism is especially needed in case of modern multi-purpose IoT gateways, where both demand and operation conditions can vary largely between deployments. To facilitate the development and operations of gateways, we implement offloading directly as part of the IoT rapid prototyping process embedded in the software stack, based on Node-RED. We evaluate the implemented method using an image processing example, and compare various offloading strategies based on resource consumption and other system metrics, highlighting the differences in handling demand and service levels reached.
\end{abstract}


\section{Introduction}

Fog computing enables use cases where data produced in end devices are stored, processed, and acted on close to the edges of the network, preceding,  augmenting or even avoiding processing in the distant cloud and the data-centers associated with it~\cite{Bonomi2014}. For fog computing to flourish, however, cloud-like simplicity should be achieved in the development and the operations (DevOps) of these systems and services, a goal that only recently started to be addressed by the community with the use of containerization~\cite{Bellavista:2017:FFC:3007748.3007777,Hoque8029944,globecom2018}.

In the case of IoT (Internet of Things) systems, computational and storage resources are naturally available in the gateway segment of the IoT architecture, residing between low-power IoT devices and cloud services. Recently, several industrial and community-based initiatives focused on the development of software stacks for such gateways aiming to simplify DevOps for IoT services running on state-of-the-art IoT gateway hardware; providing readily available software components, rapid prototyping tools, cloud connectors, and device-, gateway-, fleet-management services. Although these platforms have scarce resources compared to servers in data-centers, modern hardware allows them to run complex processing tasks by deploying applications on the edge. Scalability is however still a challenge due to the limited resources and often widely varying operating conditions. 

In this paper we present a practical approach to mitigate scalability problems by dynamically offloading computational tasks from edge to cloud resources, as an embedded part of a rapid prototyping environment, thus simplifying both the development and operation of such services.

Our paper provides the following contributions to the literature on offloading from edge to cloud:
\begin{itemize}
\item we introduce the concept of the on-demand offloading of computational tasks from IoT gateways to cloud instances and provide an implementation in Node-RED;
\item we identify offloading strategies and quantify some of the trade-offs involved.
\end{itemize}

The paper is structured as follows: after a brief introduction to the state-of-the-art in Sec.~\ref{sec:soa}, we introduce the role of rapid prototyping in the IoT gateway stack in Sec.~\ref{sec:rapid} and \ref{sec:nodered}. Our implementation of offloading is introduced in Sec.~\ref{sec:offloading} and evaluated through an image processing example in Sec.~\ref{sec:validation}.

\section{State of the Art}
\label{sec:soa}

Offloading computation from the gateway to the cloud resembles another scenario which has received considerable attention: offloading computation from mobile devices to the cloud. Enzai and Tang~\cite{6834942} present a taxonomy of the computation offloading approaches in Mobile Cloud Computing. Bangui et. al~\cite{DBLP:journals/jsan/BanguiGBRRP17} and Wu~\cite{8252700} published surveys on multi-objective decision-making to offload computation from mobile devices to the cloud. Although these approaches relate to our work, mobile devices are subject to constraints that do not necessarily apply to IoT gateways. For example, mobile devices are connected to networks with limited quotas or higher costs, and although this can also be the case for IoT gateways, it does not necessarily have to be the case. A similar consideration applies when considering energy consumption.

In the IoT gateways domain, pub/sub solutions have been considered to achieve computing offloading. Happ et al. ~\cite{Happ2017} analyze the main requirements of IoT platforms and evaluate how some existing pub/sub meet those requirements and their performance. Happ and Wolisz have described an approach to offload computation from IoT gateways to the cloud, for pub/sub services~\cite{7946415}. Specifically, they propose to develop pub/sub services that expose interfaces to set and retrieve the current state of the service, along with other functions, to let a hypervisor do the real-time migration of services from the gateway to the cloud. Even though the approach presented by Happ and Wolisz relates to our concept because it offloads computation to the gateway, our approach considers different conditions. On the one hand, our approach leverages the programming paradigm of Node-RED to simplify offloading computation. On the other hand, we offload complete jobs, thus making the process of offloading simple and consistent, i.e., no issues with synchronization of states.

\section{Rapid prototyping in the IoT gateway stack}
\label{sec:rapid}

Prototyping and development of IoT solutions have recently become accessible thanks to the availability of open hardware, software, and development kits for implementing devices (the Things in the IoT). Solutions also matured on the cloud side with PaaS resources and IoT specific cloud services readily available on the market.
Between the two, at the IoT gateway level, single board computer platforms (such as the Raspberry Pi family of boards) simplified development, but the complexity remained at the software side.

To address this lack of software frameworks at the gateway level, several initiatives started development of software stacks for IoT gateways, such as Eclipse Kura focusing on industrial use cases, Eclipse SmartHome targeting home automation, or the Linux Foundation's EdgeX Foundry focusing on the development of hardware and OS-agnostic framework for edge computing. Similar to these, AGILE also develops a software stack for IoT gateways, but its focus is to empower the gateway segment of the IoT architecture by utilizing its storage and computation resources and at the same time, to address challenges in rapid-prototyping of IoT applications.

The AGILE gateway software architecture supports the development, deployment, and execution of IoT applications directly on the gateway and facilitates the connection of these applications to things southbound and to cloud services northbound. AGILE provides a complete software stack for IoT gateways, supporting two main types of applications: i) full-fledged applications can be built using SDKs and installed on the gateway as Docker images; ii) rapid prototyping of applications using event-based visual code composition is embedded in the framework. In this paper, we focus on this second case and show how execution off-loading can be seamlessly integrated into this rapid-prototyping process to enable applications that can extend to cloud computing resources regardless of the underlying platform.

\section{Event-Based Visual Code Composition}
\label{sec:nodered}

\begin{figure}
\centering
\includegraphics[width=0.4\textwidth, trim = 0cm 0cm 0cm 0cm, clip=true]{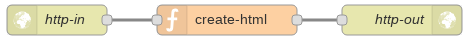}
\caption{Http server implemented as a Node-RED flow}
\label{fig:flow}
\end{figure}

Node-RED~\cite{nodered} provides an intuitive graphical interface to event-driven programs, also called  ``flows''. A Node-RED flow consists of several nodes interconnected by wires, where each node contains configurations, credentials or code. Messages are triggered by input nodes, e.g., sensors, and are passed and processed along the flow. Figure~\ref{fig:flow} depicts a simple Web server implemented in Node-RED. The first node ensures that when the flow is deployed, the Node-RED instance executing the flow listens for a particular URL. The node in the middle is a function node including the necessary JavaScript code to create a sample HTML page. The last node sends the HTTP response corresponding to the HTTP request triggering the node execution. 
In addition to its simplicity, Node-RED offers high flexibility through a wide range of nodes for different abstractions and network layers: HTTP, MQTT, TCP, UDP...

Another interesting aspect of Node-RED is that it allows developers to separate their business logic into separate flows. In this way, users can split their code into different modules that interact with each other once they are deployed. This property is of utmost importance for our approach because in this paper we explore the impact of offloading particular flows to the cloud while keeping a simple ``shell'' flow running locally. In turn, the local flow interacts with flows deployed in the cloud taking care of the heavier computation.

\section{Dynamic Offloading of Computation}
\label{sec:offloading}

The computational capacity of IoT gateways can largely vary based on the hardware platform of choice, ranging from single-core passively cooled ARMv7 platforms with 512 MB of RAM to multi-core platforms with AMD64 processors and several GBs of RAM, and recently, with different GPU-based, image processing, or AI specific accelerators. When gateways are used for multiple purposes, the share of resources available to a single application can also vary depending on the long-term and instantaneous load generated by other services. Moreover, environmental factors, such as external temperature can also make computational capacity much more variable due to thermal limitations than what one can count on in temperature controlled data centers. These limits, however, also depend on the actual hardware configuration (e.g., extension shields), on the casing applied, or, in the case of fast prototyping, even on the quality of assembly.

At the same time, applications requiring relatively heavy computation are getting more traction in several verticals of IoT. E.g., home automation needs face and image recognition, industrial automation uses anomaly detection for the manufacturing process,  machinery needs predictive maintenance.  

Our goal is to support the rapid-prototyping of such applications directly in Node-RED as part of the AGILE platform, and at the same time make sure that solutions are robust to variations in application load and operation conditions.  

\paragraph{The Offloading Mechanism}
\label{sec:offloading-mechanism}

To support offloading as an integral part of rapid application development, we developed Node-RED extensions that facilitate programming and system deployment. At Node-RED level, the developer should first develop his application without dealing with offloading. The only change required is to identify the computationally intensive part of the application logic, by embedding it in a sub-flow, connected with so-called link nodes to the main flow. Once the flow is deployed, the selected part of the workflow is also automatically deployed on other computation nodes, and these remote instances get automatically connected to the original workflow. Therefore, we automatically get a modified workflow that enables both local and remote execution of the computationally intensive part. The modified workflow is visualized, and it is further tunable by the developer.  For this, we have developed two extensions to Node-RED:
\begin{itemize}
\item TabDeployer: this takes Node-RED flows as input, and automatically deploys selected parts of them in other Node-RED instances. The functionality is exposed in the development user interface, but it can also be triggered automatically. While deploying the remote flow, TabDeployer also rewrites the local flow to connect to the remote instance using the CloudLink node.
\item CloudLink: this node is responsible for redirecting execution to remote nodes if and when needed. It contains all the decision logic to determine whether execution could remain local or whether parameters should be sent to the remote instance. The node is also responsible for handling the communication with the remote Node-RED instance, hiding all the complexity from the developer. It currently uses HTTP for communication, while pub/sub (MQTT and Cote) support is being developed. 
\end{itemize}

\paragraph{Making Offloading Dynamic}
\label{sec:offloading-dynamic}

In this paper, we focus on use cases where the scope of offloading is to guarantee system operation in case of sudden load peaks or other overload situations, e.g. due to background processes on the gateway. 
In other words, offloading should only happen when local computational resources are scarce. When such offloading happens, service performance could be compromised due to communication delays, communication costs might occur, and cloud resources are being used that should be paid. However, execution goes on seamlessly, and temporary, as well as long-term, overload situations could be handled without service interruption. To achieve this, we have identified the following resources that are monitored, and which are used to make a local decision whether to offload the computation or not:
\begin{itemize}
\item memory utilization: the availability of memory resources on the common gateway platforms can be seen as a hard limit. Swapping to disk is mostly out of the question due to slow I/O leading to further overload, and thus should be avoided at all cost. 
\item instantaneous CPU utilization: CPU utilization is an important indication that completion time will be compromised. However, its instantaneous version is susceptible to sudden peaks due to the operation of other background processes; therefore, it should be used with care, e.g., \ in combination with other metrics. CPU utilization can also be misleading due to dynamic frequency scaling (throttling) applied in under-load situations.
\item CPU temperature: temperature is also a good metric of longer-term CPU activity. Moreover, since temperature is also used to keep the platform's operation stable by throttling (reducing the speed of) the CPU, it can be used to keep the CPU at optimal operating conditions.
\item number of computation tasks currently in execution locally: finally, limiting the number of tasks executed locally can provide simple means to ensure that the local system is not overloaded. Tasks over the threshold could be delayed or offloaded, of which we study the latter case.        
\end{itemize}

\section{Usage Example and Experimental Validation}
\label{sec:validation}

We study the offloading mechanism and decision rules on an example application based on image recognition, where we use the Tesseract library to extract text from images\footnote{All the code used for the experiments is available at https://github.com/Agile-IoT/nodered-rpi-offload/tree/cloudcom}. Whenever a new image is pushed to the application\footnote{to simplify the measurements we always use the same image}, processing starts and all text contained in the image is extracted. 

We validate the offloading mechanism by first measuring the operational characteristics of a typical gateway platform under load when offloading is not enabled. Then, offloading is deployed, and we compare different off-loading strategies.

\paragraph{Gateway Performance Without Offloading}

Our experimental setup is composed of a passively cooled Raspberry Pi 3 model B with a quad-core ARM processor, 1GB of RAM, and a small heatsink, connected to an OpenStack cloud instance prepared to run the offloaded code segments.

\begin{figure}
\centering
\includegraphics[width=\columnwidth, trim = 0cm 0cm 0cm 0cm, clip=true]{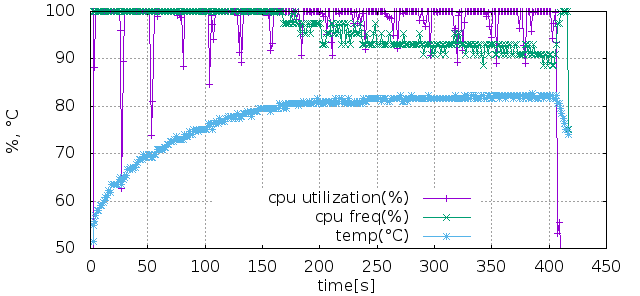}
\includegraphics[width=\columnwidth, trim = 0cm 0cm 0cm 0cm, clip=true]{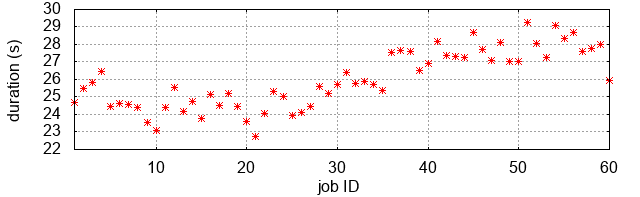}
\caption{Performance on Raspberry Pi without offloading}
\label{fig:pi}
\end{figure}

First, we characterize the performance of the gateway by running the application with constant load making sure 4 computationally intensive tasks (from now on jobs) are always running in parallel, i.e., starting 4 jobs and injecting a new job as soon as one is finished. During execution, we measure system metrics (memory utilization, CPU utilization, CPU temperature, CPU frequency), and job metrics (completion time, success ratio) in parallel. Fig.~\ref{fig:pi} shows the most important metrics as a function of time (top) and job ID (bottom).

As expected, CPU utilization is 100\% almost all the time as all four cores are loaded with jobs, and lower values are only present while jobs are resubmitted. Note how execution changes as the gateway heats up reaching limit temperatures where the system restricts CPU performance. In the first 170~s of execution, job duration fluctuates between 23 and 26~s due to other (random) background tasks in the OS. Performance degrades notably after about 170 seconds when the CPU reaches its limit temperature (80 \degree C). As CPU frequency is scaled down by the system, job durations gradually reach 29~s.

\paragraph{Offloading in Action}

\begin{table}[htbp]
 \centering
 \caption{performance with different offloading strategies}
 \label{tab:pi-perf}
 \setlength\tabcolsep{5pt} %
 \begin{tabular}{|l|l|l|p{.6cm}|    p{.94cm}|p{.94cm}|}
  \hline
  Name & Metric & Limit  & Local jobs (\%)  & Average duration (s) & Max. duration (s) \\
  \hline
  $Jobs_{4}$ & Jobs in execution & 4 & 70.0 & 26.5 & 31.7 \\
  \hline
  $CPU_{75\%}$ & CPU utilization & 75~\% & 62.5 & 19.1 & 20.7 \\
  \hline
  $Mem_{75\%}$ & Mem. utilization & 75~\% & 70.8 & 37.3 & 48.1 \\
  \hline
  $Temp_{75\degree}$ & Temperature & 75~{\degree}C & 53.3  & 20.5 & 39.2 \\
  \hline
 \end{tabular}
 \end{table}

\begin{figure}
\centering
\includegraphics[width=\columnwidth, trim = 0cm 0cm 0cm 0cm, clip=true]{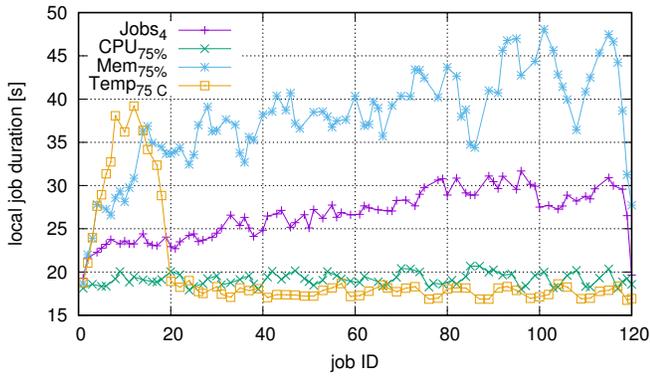}
\caption{Local execution times with different offloading strategies. Dots connected with lines only to enhance visibility.}
\label{fig:pi-duration}
\end{figure}

In the following experiments, we compare different offloading strategies. We set up the system to receive a new task every 5 seconds (to simplify comparison we use constant inter-arrival times) and configure four different offloading strategies limiting i) the number of jobs running locally in parallel; ii) CPU utilization; iii) memory utilization; iv) CPU temperature. All other jobs are offloaded to the cloud. Table~\ref{tab:pi-perf} contains selected threshold parameters and statistics about these experiments, while Fig.~\ref{fig:pi-duration} compares the time series of job execution times in detail, explaining the rationale behind the statistics.

Starting our analysis from the ratio of locally executed jobs, which in turn determines how much cloud and communication resources should be paid for, there are clear differences between the strategies with the selected thresholds. $Jobs_{4}$ and $CPU_{75\%}$ reached 70\% local execution, while other strategies offloaded more jobs to the cloud. From the statistics, the execution times of local jobs (both average and max) seems to under control with the $CPU_{75\%}$ strategy. 

Although the above statistics are interesting, the detailed Fig.~\ref{fig:pi-duration} explains more about the different behavior of these strategies. 
Limiting the number of local jobs ($Jobs_{4}$) manages to limit execution times somewhat, although it does grow as the system heats up, reaching 30~s. $CPU_{75\%}$ behaves much better in this regard, although note that it only executed 62.5\% of jobs locally. $Mem_{75\%}$ behaves similar to ($Jobs_{4}$), reaching slightly higher offloading but also higher execution times. Interesting to note the periodicity of the control, more or less evident on each figure with different periods\footnote{we use time series from a single experiment here to provide more insight}. Temperature is regarded by many as an early indicator of system stress, and indeed $Temp_{75\degree}$ provides an excellent offloading strategy, apart from an initial transient where too many jobs were allowed in local execution and thus execution times extended.

\section{Conclusions}    
\label{sec:conclusion}
    
Although single board computers are getting more and more powerful, offloading computation from IoT gateways to the cloud is necessary in several use cases and under various operating conditions to provide robust solutions at the edge. We have integrated offloading of computation tasks as part of our rapid prototyping workflow, and shown how it can dynamically handle demand keeping the gateway under load but offloading excess tasks. Our analysis of various offloading criteria revealed interesting differences in the service levels each one could provide. An extended study of multi-criteria decisions to provide service level agreement guarantees is still future work.

\bibliographystyle{IEEEtran}
\bibliography{juan} 

\end{document}